\documentclass[12pt]{article}
\pdfoutput=1
\usepackage{jheppub}
\usepackage{tabu}
\newcommand{\be}{\begin{equation}}
\newcommand{\ee}{\end{equation}}
\newcommand{\lan}{\langle}
\newcommand{\ran}{\rangle}

\begin{document}

\title{Inflation After False Vacuum Decay: \\ New Evidence From BICEP2}

\author[a,b]{Raphael Bousso,}
\author[c]{Daniel Harlow,}
\author[d,e]{and Leonardo Senatore}
\affiliation[a]{Center for Theoretical Physics and Department of Physics,\\
 University of California, Berkeley, CA 94720, U.S.A.}
\affiliation[b]{Lawrence Berkeley National Laboratory, Berkeley, CA 94720,
  U.S.A.}  
\affiliation[c]{Princeton Center for Theoretical Science, Princeton University, Princeton NJ 08540 USA} 
\affiliation[d]{Stanford Institute for Theoretical Physics,
 Stanford University, Stanford, CA 94306} 
\affiliation[e]{Kavli Institute for Particle Astrophysics and Cosmology, \\
Stanford University and SLAC, Menlo Park, CA 94025} 

\emailAdd{bousso@lbl.gov}
\emailAdd{dharlow@princeton.edu}
\emailAdd{senatore@stanford.edu} 

\abstract{Last year we argued that if slow-roll inflation followed the decay of a false vacuum in a large landscape, the steepening of the scalar potential between the inflationary plateau and the barrier generically leads to a potentially observable suppression of the scalar power spectrum  at large distances.  Here we revisit this analysis in light of the recent BICEP2 results.  Assuming that both the BICEP2 B-mode signal and the \textsl{Planck} analysis of temperature fluctuations hold up, we find that the data now discriminate more sharply between our scenario and $\Lambda$CDM. Nonzero tensor modes exclude standard $\Lambda$CDM with notable but not yet conclusive confidence: at $\sim 3.8\,\sigma$ if $r=0.2$, or at $\sim 3.5\,\sigma$ if $r=0.15$.  Of the two steepening models of our previous work, one is now ruled out by existing bounds on spatial curvature.  The other entirely reconciles the tension between BICEP2 and Planck.  Upcoming $EE$ polarization measurements have the potential to rule out unmodified $\Lambda$CDM decisively. Next generation Large Scale Structure surveys can further increase the significance. More precise measurements of $BB$ at low $\ell$ will help distinguish our scenario from other  explanations. If steepening is confirmed, the prospects for detecting open curvature increase but need not be large.}

\maketitle
\section{Introduction and Summary} 
The smallness of the cosmological constant has led to the consideration of cosmological models with a large number of metastable vacua, in which our universe would arise from the decay of a false vacuum~\cite{Linde:1984ir,Banks:1984cw,Weinberg:1987dv,Bousso:2000xa,Kachru:2003aw,Susskind:2003kw}. Freivogel, Kleban, Martinez, and Susskind~\cite{Freivogel:2005vv} pointed out that in this setting, one expects the inflaton potential to steepen as it interpolates between the slow-roll plateau, where the structure and flatness of our current universe was generated, and the high potential barrier separating it from our parent vacuum, and suggested that this might lead to some observable effect in the power spectrum.  

Last year \cite{Bousso:2013uia}, we showed that in slow-roll inflation, steepening produces a very specific signal, a suppression (never an enhancement) of the scalar power at large scales.  We noted that this effect can resolve the $2-2.5 \,\sigma$ tension at low $\ell$ in the measurements of $C_{\ell}^{TT}$ by the \textsl{Planck} satellite. Since the \textsl{Planck} anomaly was too weak to provide substantial evidence for our signal, we stressed that our analysis should be regarded as a prediction for future observations. We pointed out that in this scenario a similar power suppression should \textit{not} affect the tensor spectrum in the event that it is observed, and we noted that E-mode polarization data, large scale structure, and a nonzero value of the tensor-to-scalar ratio $r$ all have the potential to enhance the significance of a lack of scalar power at low $\ell$.

Recently the BICEP2 experiment has reported a detection of primordial tensor modes, with $r=0.2^{+.07}_{-.05}$ \cite{Ade:2014xna,Ade:2014gua}.  The importance of this discovery for early universe cosmology, if confirmed by other experiments in the upcoming year or two, is difficult to overstate.  In addition to providing almost incontrovertible evidence for an early inflationary phase of our universe, for the first time we will have direct experimental evidence of physics at energies of order $1\%$ of the Planck scale (or not too far from that if different mechanisms to produce gravity waves are involved~\cite{Senatore:2011sp}).  

The purpose of this note is to reconsider the main observational aspects of \cite{Bousso:2013uia} in light of the possibility of such large values of $r$. (For theoretical motivation and a discussion of priors we refer the reader to \cite{Bousso:2013uia}.)  We will see that a value of $r$ as large as that reported by BICEP2 considerably enhances the significance of the low $\ell$ anomaly, to $3.8-4.0\,\sigma$ if $r=0.2$ and $3.5-3.7\,\sigma$ if $r=0.15$.  There has been much recent discussion over the apparent tension between large values of $r$ and the \textsl{Planck} bound of $r<0.11$ \cite{Ade:2013uln}. An appealing theoretical interpretation is that the tension lies between the two experiments taken together and $\Lambda$CDM$+r$, rather than between the two experiments.  Obviously this could change if the experimental values for the parameters do not stay where they are.  If both the BICEP2 and the \textsl{Planck} results hold up to further scrutiny, then it now seems quite likely that the scalar primordial power spectrum deviates from $\Lambda$CDM in a specific way---suppression at large scales--- that arises rather naturally in a cosmology with many vacua.\footnote{In the BICEP2 release it was suggested that modifying $\Lambda$CDM by including running of the tilt can lessen the tension with \textsl{Planck}.  This is somewhat true, but the large running required is inconsistent with simple slow-roll models.  More concretely, what is needed to make running work is for the potential to have a surprisingly large third derivative in the vicinity of the value of $\phi$ corresponding to $\ell=700$. The effect needed is so large that it would make inflation not last long enough to get a sufficient number of e-foldings before reheating, unless we also introduce a drastic re-flattening of the potential at $\ell\gg 700$~\cite{Easther:2006tv}. 
We see no theoretical motivation whatsoever for such a feature, unlike the potential steepening at low $\ell$ that we are considering in this paper. Regardless of this theoretical bias, running does not fit the data as easily as a steepening feature. This is because steepening improves the fit precisely in the low $\ell$ region where there is tension with $\Lambda$CDM, whereas running  creates new tension at high $\ell$.}

Among the many papers that followed the announcement of the BICEP2 results while this work was in progress, some overlap with ours, see for example~\cite{Miranda:2014wga,Smith:2014kka,Hazra:2014aea,Hazra:2014jka}. In particular, the connection to our earlier work \cite{Bousso:2013uia} was noticed in \cite{Hazra:2014aea,Hazra:2014jka}.  Earlier related work includes \cite{Linde:1998iw,Garriga:1998he,Linde:1999wv,Contaldi:2003zv,Yamauchi:2011qq,Liddle:2013czu,Dudas:2012vv,Pedro:2013pba}.

\section{Tensor Contribution to the Temperature Anisotropy}
The origin of the increased significance of the observed lack of scalar power is that primordial tensor modes indirectly enhance $C_{\ell}^{TT}$ at low $\ell$, but are damped above $\ell\approx100$ \cite{Weinberg:2008zzc}.  The basic physics of this is straightforward to understand. Tensor modes do not couple to matter with a perfect fluid equation of state, so to a good approximation, each independent helicity mode $\phi_{\pm,q}$ of a gravitational wave in an FRW background obeys a free scalar equation\footnote{Our conventions here are the same as in \cite{Bousso:2013uia}; $a$ is the dimensionful scale factor and $q$ is the dimensionless comoving wave number.  The dimensionful wave number is $k=q/a_0$.}
\be
\ddot{\phi}_{\pm,q}+3H\dot{\phi}_{\pm,q}+\frac{q^2}{a^2}\phi_{\pm,q}=0.
\ee
The relative importance of the second and third terms depends on the size of $\frac{q}{a H}$.  If this quantity is small then the mode is outside the horizon and we can neglect the third term, so the solution quickly approaches a constant.  If this quantity is large, the mode is inside the horizon and second term is subleading.  The mode then undergoes a damped oscillation, decaying as $1/a$.  At reheating all interesting tensor modes are far outside of the horizon, with their constant value set by the primordial distribution generated during inflation.  During radiation and matter domination they gradually re-enter and damp away.  Only tensor modes which entered the horizon not much before last scattering can lead to important effects in the CMB.\footnote{This argument does not so directly apply to scalar perturbations, since these are substantially sourced by matter and radiation and thus obey more complicated equations after horizon entry.}  These are the modes with $\ell\lesssim 100$.

\begin{figure}
\begin{center}
\includegraphics[height=6cm]{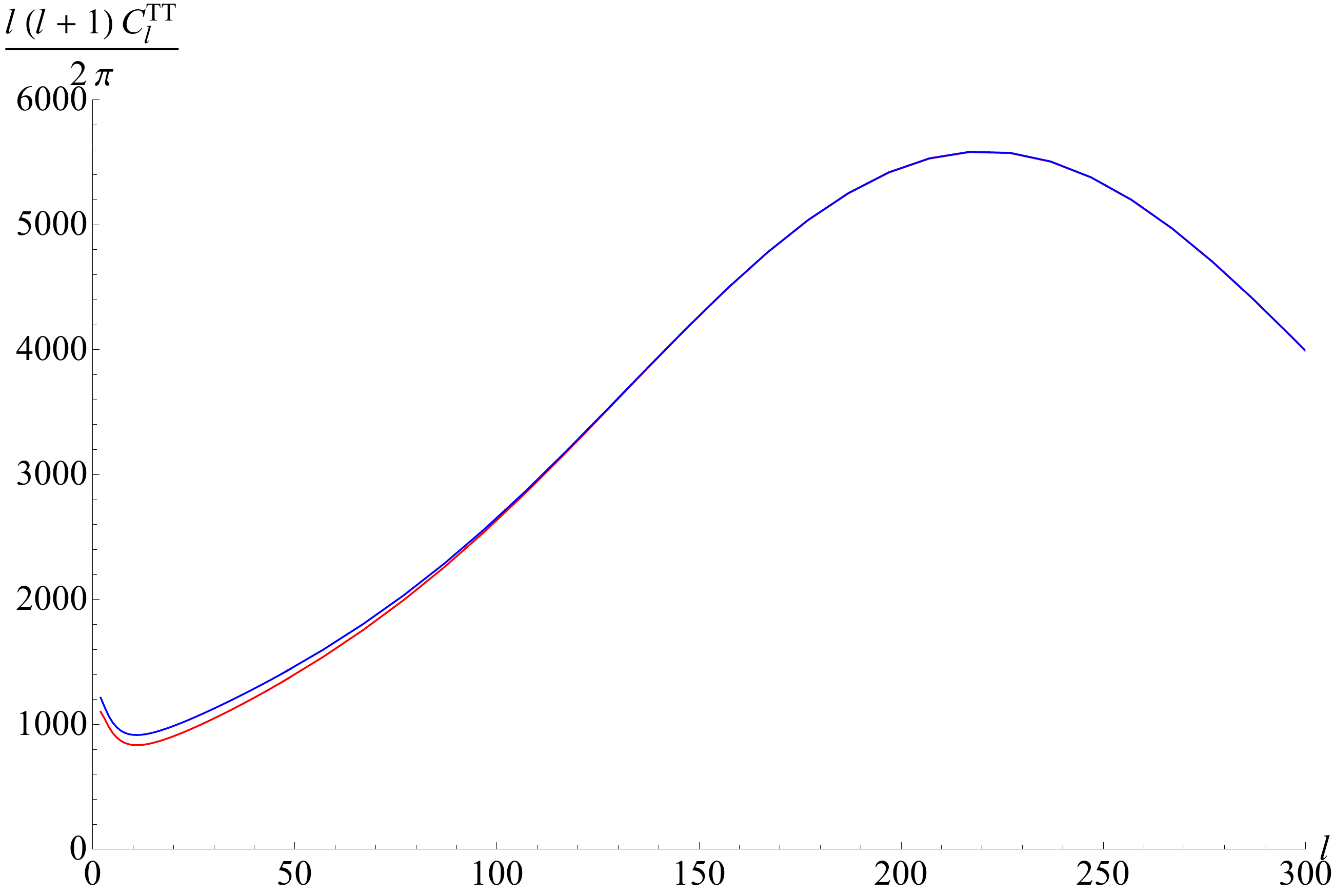}
\caption{$\Lambda$CDM predictions for $C_\ell^{TT}$ with the \textsl{Planck} parameters, with $r=0$ in red and with $r=0.2$ in blue.  As expected, the difference between the two curves dies off above $\ell\approx 60$.}\label{tensorsupp}
\end{center}
\end{figure}

The contribution from primordial gravitational waves to the CMB anisotropy can easily be discerned in figure \ref{tensorsupp}, which shows $C_\ell^{TT}$ evaluated for $r=0$ and $r=0.2$ $\Lambda$CDM as calculated using CLASS \cite{Lesgourgues:2011re,Blas:2011rf}.
  
Figure \ref{tensorsupp} is very illuminating from the point of view of the discrepancy between the Planck bound of $r<0.11$ and the BICEP2 detection of $r\approx 0.2$.  As we will review momentarily the \textsl{Planck} scalar CMB spectrum lies somewhat \textit{below} the $r=0$ red curve ($\Lambda$CDM, r=0) at low $\ell$. Any modification of the theory that pushes the predicted curve up, such as $r>0$, is thus disfavored. This is how \textsl{Planck} was able to bound $r$ without using B-mode polarization measurements.  

The implicit assumption, however, is that $\Lambda$CDM is unmodified at small $\ell$. If the contribution to $C_\ell^{TT}$  from scalar fluctuations is substantially suppressed due to a feature in the inflaton potential, then the mildly low $C_\ell^{TT}$ power seen by \textsl{Planck} could contain a sizable contribution from primordial graviational waves. Conversely if primordial tensor modes are observed directly through the B-mode spectrum of the CMB, as claimed by BICEP2, then this greatly increases our confidence that the primordial scalar spectrum is indeed suppressed at $\ell$ below $30-50$. 

\section{Increased Significance of the Anomaly}\label{sigsec}
We now quantify this increased confidence, using the crude statistical methods we introduced in \cite{Bousso:2013uia}. 
\begin{figure}
\begin{center}
\includegraphics[height=8cm]{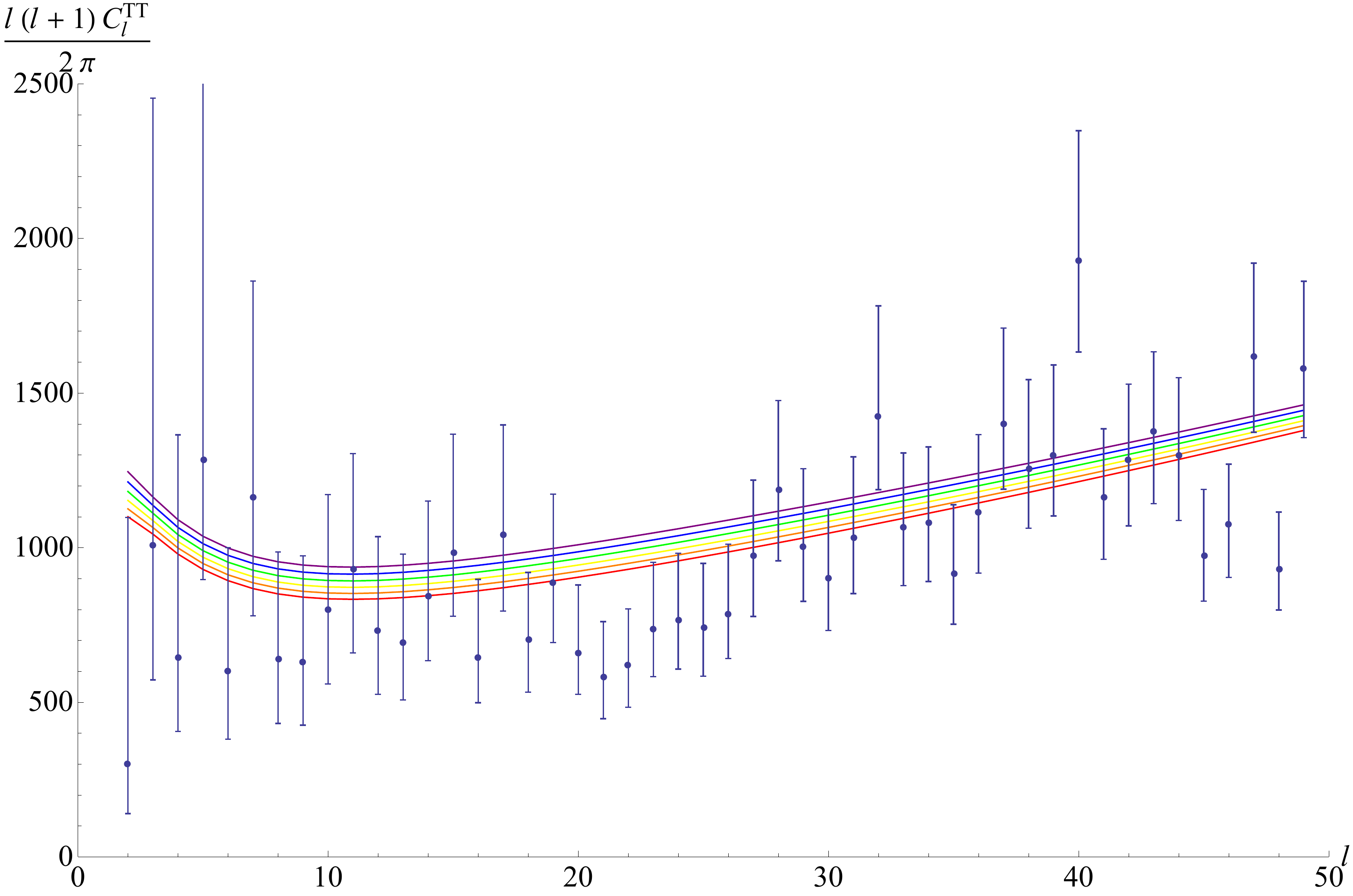}
\caption{$\Lambda$CDM predictions for $C_\ell^{TT}$ at low $\ell$, for $r=0$, $0.05$, $0.1$, $0.15$, $0.2$, and $0.25$.  The data are from \textsl{Planck} \cite{Ade:2013kta}, and the theory curves are computed using CLASS with the \textsl{Planck} best-fit values for the parameters of $\Lambda$CDM.}\label{rainbow}
\end{center}
\end{figure}
\begin{table}
\begin{center}
\begin{tabular}{|l | l |l |}
\hline
r & $\ell_{max}=30$ & $\ell_{max}=50$ \\
\hline
0 & 2.4 & 1.7\\
0.1 & 3.0 & 2.7\\
0.15 & 3.7 & 3.5 \\
0.2 & 4.0 & 3.8\\
0.25 &4.2 & 4.4\\
\hline
\end{tabular}
\caption{$\sigma$ values for two different estimators quantifying the deviation of the data from the theory curves in figure \ref{rainbow}.}\label{tabsig}
\end{center}
\end{table}
Figure \ref{rainbow} shows the growing tension between the data and $\Lambda$CDM+$r$ as the tensor contribution to $C_\ell^{TT}$ is increased.  Table \ref{tabsig} contains the results of our significance analysis.  The two columns correspond to different estimators. The first looks for primordial power suppression below $\ell=30$; the second looks for suppression below $\ell=50$.  The values at $r=0$ are consistent with the results of the \textsl{Planck} collaboration, who found a significance for the suppression of order $2-2.5 \,\sigma$ \cite{Ade:2013kta}. Importantly, we find that the contribution of the primordial tensor modes enhances the significance considerably.\footnote{In doing this analysis, we have used the cosmological parameters reported by the Planck collaboration \cite{Ade:2013zuv}.  In principle we should allow them to vary to counter the surprisingly large value of $r$, but this will not be of much help since changing them significantly would disrupt the beautiful fit at higher $\ell$.  Recently however Spergel, Flauger, and Hlozek have re-analyzed the data using a different cleaning strategy for the maps, and found interesting parameter shifts \cite{Spergel:2013rxa} which go in the direction of returning the parameters to their WMAP9 values \cite{Hinshaw:2012aka}.  This would end up decreasing our significance values somewhat, since after all the increased tension at low $\ell$ reported by \textsl{Planck} was mostly a consequence of the parameter shifts from WMAP9.  We are in no position to take a side in this debate, but we hope it will be resolved in the upcoming \textsl{Planck} data release.} Even though the techniques we have used here are crude, we expect that more accurate methods will not shift the significances by very much, and therefore we consider our methods sufficient to illustrate the point.   
 
\section{Models}\label{modsec}
In $\Lambda$CDM one assumes a standard form of the primordial scalar power spectrum characterized by an amplitude and a tilt:
\be\label{stP}
k^3 P_{scalar}(k)=2\pi^2 A_s\left(\frac{k}{k_*}\right)^{n_s-1}~,
\ee
where $k_*$ is some reference wave number typically taken to be $0.05\, \mathrm{Mpc}^{-1}$.  $r$ is defined as the ratio of the tensor to the scalar power spectrum at $k_*$.  In \cite{Bousso:2013uia} we considered single-field inflationary models with potentials of the form
\be\label{Vdc}
V(\phi)=V_S(\phi)+\gamma V_R(\phi),
\ee
where $V_S$ is a monotonically decreasing slow-roll inflationary potential designed to fit the observed values of $A_s$, $n_s$, and $r$, and $V_R$ is a monotonically-decreasing ``steepening perturbation'' which dies off at large positive $\phi$ but begins to grow steeply in the vicinity of $\phi=0$. We took $\phi=0$ to be the value of the field during inflation at which the comoving scale of our current horizon was exiting the inflationary horizon.  As we shall see, $(\gamma V_R'/V_S')$ is the leading combination that controls the suppression of the power spectrum; it must be small for our perturbative analysis to be valid. The potential $V_R$ itself need not be slow-roll near $\phi=0$, but $V_S$ must be, so generically we will have $\chi\equiv \frac{(V_R/V_S)}{(V'_R/V'_S)}= (\epsilon_S/\epsilon_R)^{1/2}\ll 1$ in the regime of interest.  In particular this implies that corrections in $\gamma V_R/V_S$ are subleading to corrections in $\gamma V_R'/V_S'$, as $\gamma V_R/V_S= \chi\cdot \gamma V_R'/V_S'\ll \gamma V_R'/V_S'$.

We found that to leading orders in $\gamma$ and $\chi$, one has scalar and tensor power spectra
\begin{align}
P_{scalar}&\approx P_{scalar,S}\left[1-2\gamma\frac{V_R'}{V_S'}+O\left(\chi\cdot\frac{\gamma V_R'}{V_S'},\left( \frac{\gamma V_R'}{V_S'}\right)^2\right)\right]\ , \\
P_{tensor}&\approx P_{tensor,S}\left[1+O\left(\chi\cdot \frac{\gamma V_R'}{V_S'}\right)\right]\ .
\end{align}

Hence, the scalar power has a suppression of order $\gamma V_R'/V_S'$, whereas the tensor spectrum is unchanged at this order.  The idea was that $V_S$ should be chosen to fit all data \textit{except} for the deficit of scalar CMB power seen by Planck at low $\ell$. $V_R$ should steepen the potential appropriately, to account for the lack of power at the largest scales. This is in practice an independent choice, though one must keep track of whether the perturbation invalidates the slow-roll conditions at some negative $\phi$, since this is constrained by present bounds on spatial curvature.

To fit $V_S$ to the best-fit Planck cosmological parameters ($A_s=2.2\times 10^{-9}$, $n_s=0.96$), and now also to BICEP2 ($r=0.2$), we expand $V_S$ near $\phi=0$ as 
\be
V_S(\phi)=V_i\left(1-\sqrt{2\epsilon_S}\phi+\frac{\eta_S}{2}\phi^2\right),
\ee
and choose $\epsilon_S=0.013$, $\eta_S=0.019$, and $V_i=7.9 \times 10^{-9}$ (here we have set $M_P^{-2}\equiv 8\pi G=1$).\footnote{A minor subtlety here is that $\epsilon_S$ and $\eta_S$ are defined at the horizon scale instead of $k_*$; the difference is subleading in the slow-roll parameters so we will ignore it.  Moreover although we have included nonzero $\eta_S$ in order to get the BICEP2 value of $r$, we will be able to drop it in our results for the scalar power spectrum below since we are only interested in low $\ell$, where $\phi \sim \sqrt{2\epsilon_S}$.}  

In \cite{Bousso:2013uia} we gave two toy examples of possible choices for $V_R$, which we used to fit the Planck anomaly, which we will now revisit.  Our first model, in which $V_R$ was a simple exponential, was highly constrained by limits on spatial curvature. If this model is adjusted to provide a primordial scalar suppression large enough to overcome the tensor contribution to $C_{\ell}^{TT}$ favored by BICEP2, then we find that it also predicts greater negative curvature than current bounds allow. Hence, this model is now ruled out.

Our other toy model was
\be
V_R=\theta(\phi_c-\phi) \frac{V_i}{\zeta} (\phi_c-\phi)^\zeta,
\ee
where $\theta$ is the Heaviside step function and it is convenient to replace $\phi_c$ by 
\be
\phi_c\equiv \sqrt{2\epsilon_S}\log \frac{\ell_c}{D_{LS} H_0},
\ee
where $\ell_c$ is approximately the $\ell$ in the CMB where the perturbation turns off.  To leading order in the slow roll parameters the scalar power spectrum for this model is
\be\label{modelP}
P_{scalar}(k)\approx P_{scalar,S}\left[1-\frac{\gamma}{\sqrt{2\epsilon_S}}\left(\sqrt{2\epsilon_S} \log \frac{\ell_C}{k D_{LS}}\right)^{\zeta-1}\right].
\ee
For $10\lesssim \ell\lesssim 100$ we can approximately include the effect of this correction on $C_\ell^{TT}$ as follows. We replace $k\to \frac{\ell}{D_{LS}}$ and multiply the $\Lambda$CDM result for $C_\ell^{TT}$ with $r=0$ (as computed by CLASS) by the factor in square brackets in the above equation. This gives the primordial scalar contribution to the $C_\ell^{TT}$ CMB spectrum.  To this we add the difference between the $\Lambda$CDM CLASS prediction for $r=0.2$ and $r=0$, in order to include the tensor contribution, which is unmodified at the order we are working at.  This approximation is allowed since in this range there is no evolution of the perturbations and we are essentially just seeing the primordial spectrum directly in the CMB.  We construct the modified power spectrum in this way also for $\ell\lesssim 10$. Due to the ISW effect and the breaking of the flat sky approximation, our treatment is only marginally justified for $\ell \lesssim 10$, but given the large cosmic variance of the very low $\ell$ modes, and given that we aim at performing only an approximate study of the statistical significance,  we consider this approximate treatment justified.\footnote{Julien Lesgourgues has kindly modified CLASS to include our exponential model from \cite{Bousso:2013uia}, so in that case we were able to check this approximation against a proper numerical evolution of the primordial spectrum to the CMB.  We found it to be quite accurate, so we were comfortable using it in our discussion of the model we also consider here.}

\begin{figure}
\begin{center}
\includegraphics[height=7cm]{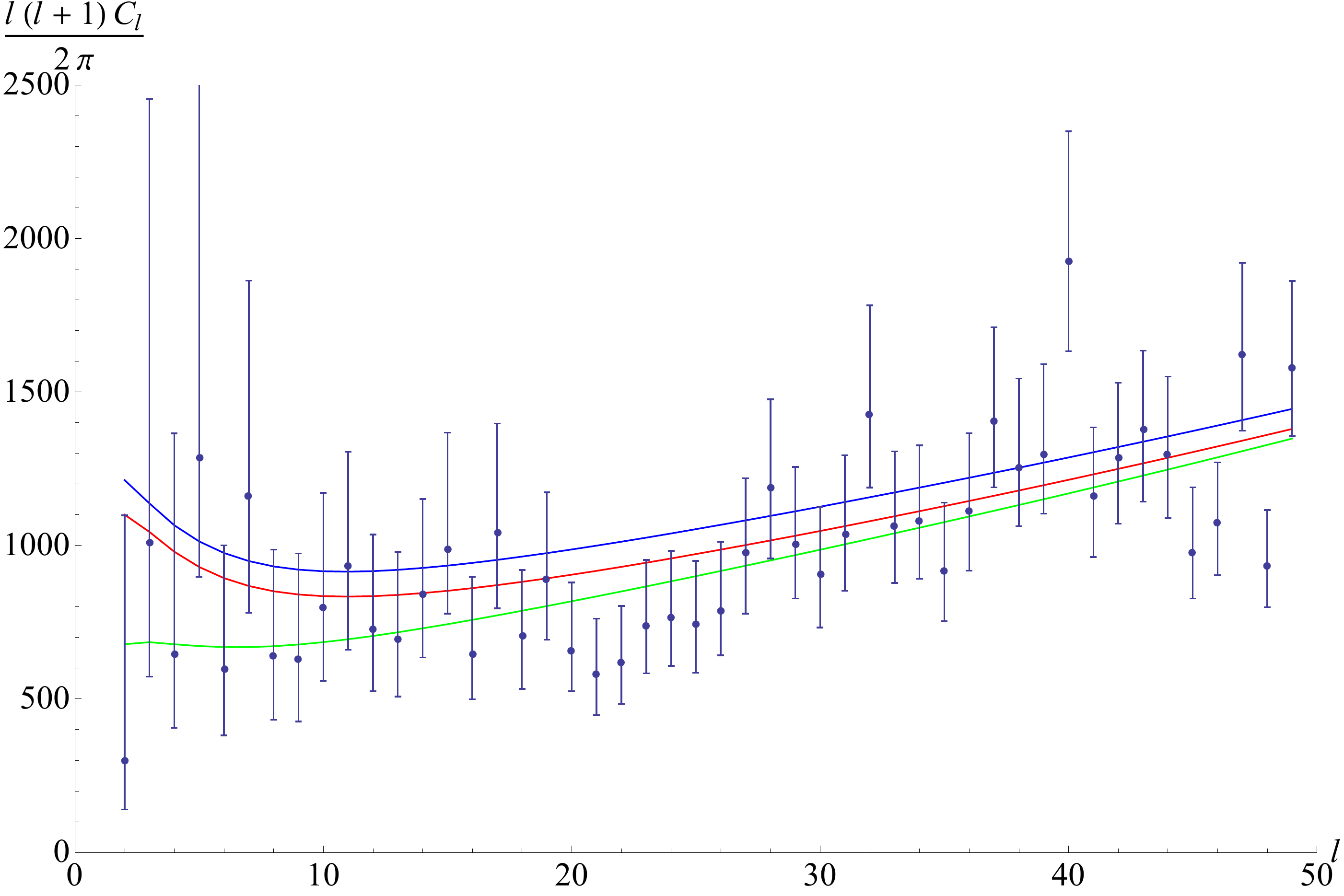}
\caption{Our model \eqref{modelP} in green, plotted against $\Lambda$CDM with $r=0$ in red and $r=0.2$ in blue.}\label{modelfit}
\end{center}
\end{figure}
In figure \ref{modelfit}, we show a plot of the low $C_\ell^{TT}$'s for this model, with $\gamma=0.065$, $\zeta=2$, and $\ell_c=84$.  According to our two estimators the deviation is now $0.6\,\sigma$ for $\ell_{max}=49$ and $0.3\,\sigma$ for $\ell_{max}=30$, so the model essentially removes all of the tension, nicely fitting both BICEP2 and \textsl{Planck}.

The reader may object to the non-analyticity of this toy potential, but it would be easy to write down analytic models that smooth out the junction at $\phi_c$ and accomplish the same task.  We close this section with a few comments on the general properties any model must have to get a good fit.  First of all, $V_R$ must fall off sufficiently fast at large positive $\phi$: the relationship between $\phi$ and $\ell$ is logarithmic, so any power-law fall off in $\phi$ would pollute the remarkable success of $\Lambda$CDM at high $\ell$.  The potential cannot grow too fast at negative $\phi$ however, since this would cause the beginning of inflation to be too recent and run into conflict with observational bound $|\Omega_K|<10^{-2}$ on the magnitude of spatial curvature, which is negative if our universe formed by the decay of a false vacuum.  With $r=0$ it is possible to have a single exponential that satisfies both criteria, but not when $r\gtrsim 0.1$.  In the model we have presented here the growth at negative $\phi$ is only power law, so as we emphasized in \cite{Bousso:2013uia} the model would remain viable even if bounds on curvature shrank to the cosmic variance limit.

\section{Future Polarization and Large Scale Structure Measurements}
It is obviously interesting to ask if future measurements can provide stronger observational support for the type of model we considered in section \ref{modsec}.  In \cite{Bousso:2013uia} we argued that polarization observations would help somewhat, but we focused on large scale structure surveys that should eventually be able to considerably increase the significance of the low-$\ell$ power suppression if it is real.  The BICEP2 discovery however has greatly increased the importance of upcoming polarization measurements for these purposes, as we now explain.  

First of all having access to B-modes of such large amplitude should allow a test of our observation from \cite{Bousso:2013uia} that potential steepening does \textit{not} lead to low-$\ell$ power suppression in the tensor spectrum at leading order in $\gamma V_R'/V_S'$ and $\chi$.  The upcoming \textsl{Planck} polarization data will be the first big step in this direction.  This may be a valuable way to distinguish potential steepening from other possible explanations of the low $C_\ell^{TT}$'s.  

\begin{figure}
\begin{center}
\includegraphics[height=5cm]{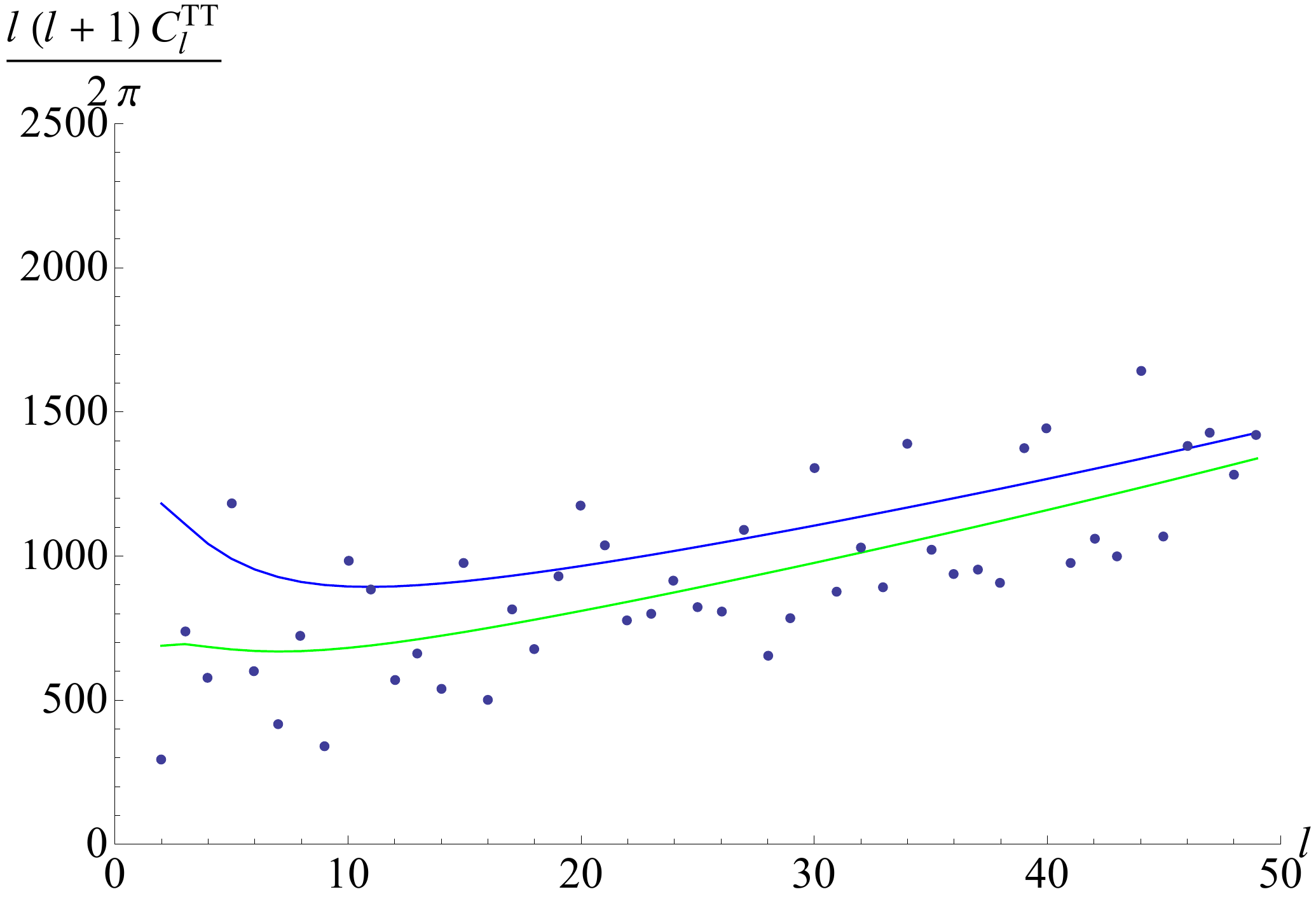}
\includegraphics[height=5cm]{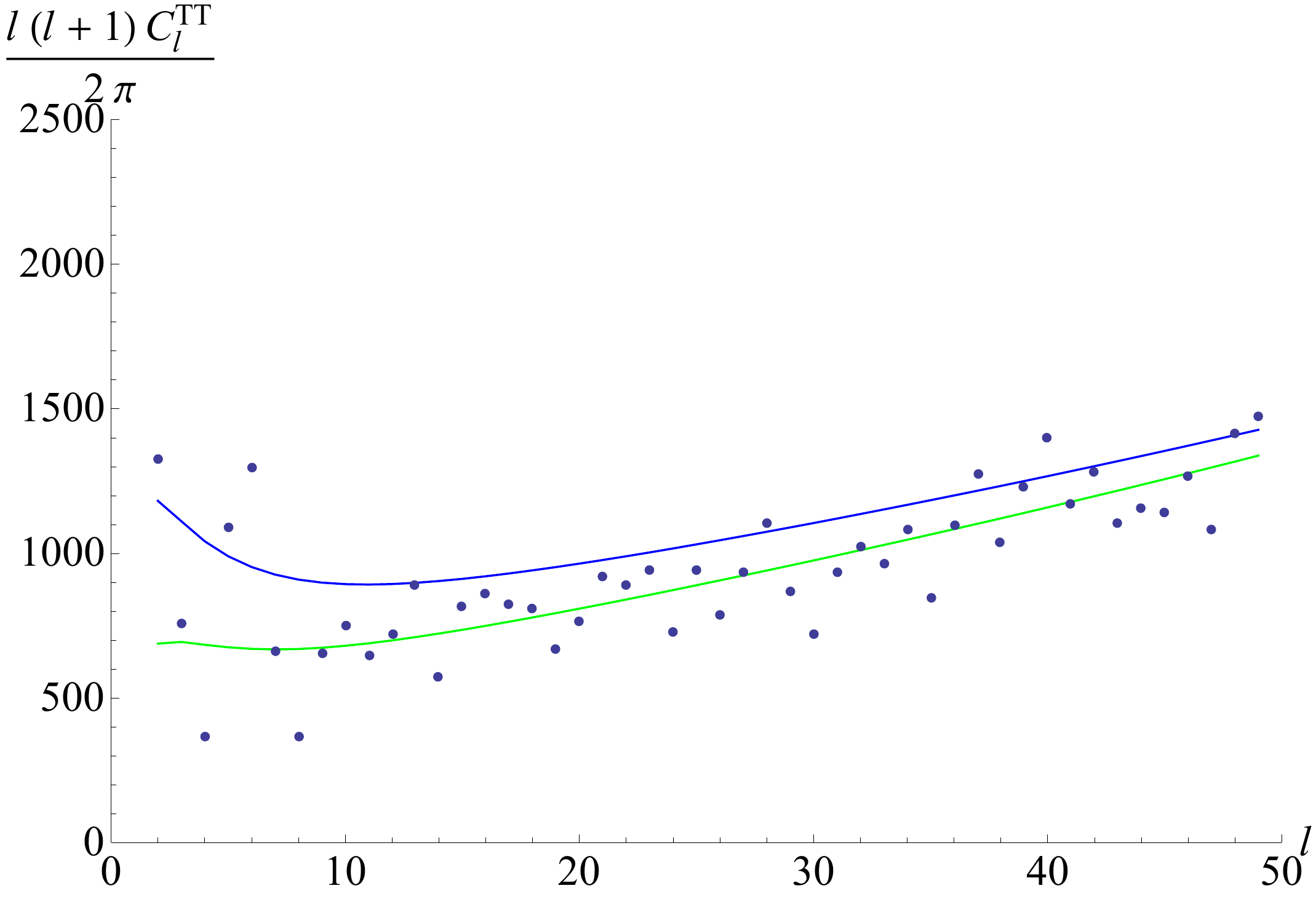}
\caption{ The left plot shows  simulated data for $C_{\ell}^{TT}$, assuming our model is correct. Using the same crude statistical method as in section \ref{sigsec}, these data deviate from the blue curve at $3.4\,\sigma$.  On the right, we illustrate the effect of the additional statistics of future $E$-mode polarization data by re-generating the temperature data using appropriately rescaled cosmic variance.  These data deviate from the blue curve at $4.9\,\sigma$.}\label{simpol}
\end{center}
\end{figure}

Secondly, as we explained in \cite{Bousso:2013uia}, low $\ell$ measurements by \textsl{Planck} of $C_{\ell}^{TE}$ and $C_{\ell}^{EE}$, which should be also suppressed at low $\ell$ in our scenario, will give us more statistics to beat down the cosmic variance error bars in the determination of the primordial scalar spectrum.  In \cite{Bousso:2013uia} we estimated that this would not provide too much of a boost in the significance, but we are now more optimistic.  The result of including polarization is roughly to multiply the $\sigma$ of the anomaly as computed in section \ref{sigsec} by a number of order $1.3-1.4$.  The significances in Table \ref{tabsig} are now considerably larger for nonzero values of $r$, so this multiplier now packs significantly more punch.  One might worry that because of the nonzero $TE$ correlation, measurements of $C_\ell^{EE}$ would not provide as much information on the primordial power spectrum as we already have from $C_\ell^{TT}$.  In fact by using measurements of $C_\ell^{TE}$, if the noise is sufficiently low, it is possible to ``decorrelate'' $T$ and $E$, after which we in principle have access to twice as many independent modes as were available just from $C_\ell^{TT}$.\footnote{We refer the reader to \cite{Bousso:2013uia} (or to any standard cosmology textbook) for more details on why more modes are important.  We explain the decorrelation procedure in a toy model in appendix \ref{decapp}.}  Since standard deviations of uncorrelated variables are added in quadrature, twice as many modes would roughly give us a multiplicative factor of $\sqrt{2}$ on the significances, which would already be able to push a $3.5\,\sigma$ deviation into $5\,\sigma$ discovery territory.  The situation is somewhat complicated by the effects of reionization at low $\ell$ \cite{Mortonson:2009qv}, but even if we only include polarization for $\ell>10$ in our mode counting the factor is still $1.37$ if we count up to $\ell=30$ and $1.41$ if we count up to $\ell=50$.  To visualize this, in figure \ref{simpol} we show the decrease in the scatter of simulated data for the power spectrum, plotted as $C_\ell^{TT}$, if we assumed that the effect of including the measurement of polarization were to rescale the variance of the $C_\ell^{TT}$. While this is not the way temperature and polarization data will actually be be packaged, it gives a nice pictorial representation of the  statistical gain that polarization data will yield.   It will clearly be very exciting to see what comes out of the \textsl{Planck} polarization analysis.

\section{Discussion}
The most robust implication of false vacuum decay is negative curvature \cite{Freivogel:2005vv}.  But unfortunately, diluting curvature is one of the fortes of inflation. We argued in \cite{Bousso:2013uia} that even if we discover a low-$\ell$ power suppression, and even if it originates from the flanks of the potential barrier separating us from our parent vacuum, we should not necessarily expect that $\Omega_K$ exceeds the cosmic variance limit of $10^{-5}$. The reason is that the slow-roll approximation is still valid in the region where the potential is just starting to steepen, so observable power suppression can arise even if there are still many e-foldings going back in time to the beginning of inflation.  Whether or not we should expect to see curvature thus depends on the rate of steepening of the potential, which is something that we do not have well-founded theoretical predictions for.\footnote{Seeing steepening does increase our \textit{chances} for detecting curvature. They may still be small, but we should certainly look for it!} 

In the absence of direct observation of negative curvature, it is tempting to declare that seeing a low-$\ell$ scalar power suppression is really just evidence for a steepening potential, not for bubble nucleation.  We are sympathetic to this point of view (after all in some positivist sense the statement is true), but we now make a few comments about this.  

Independently of whether or not there is a power suppression in the CMB, bubble nucleation has some compelling theoretical properties \cite{Bousso:2013uia}.  It extends in a controlled way our understanding of cosmology to an earlier epoch
and provides beautiful homogeneous initial conditions for inflation. Moreover, it arises naturally in a landscape setting such as that of string theory, which has the ability to explain other observations such as the smallness of the cosmological constant.  

It is thus quite reasonable to \textit{assume} that we nucleated in a bubble, and then to ask if any experimental data could give us evidence against this hypothesis.  For example, a detection of positive curvature would kill the scenario \cite{Kleban:2012ph}, but so far there is no evidence for this. In this scenario it would also be rather unnatural, however, to find any nontrivial feature in the inflationary potential \textit{other than} a steepening at early times.  This does not mean that the scenario predicts observable steepening; after all, even if it exists, the feature might be located too far up the inflaton potential, so that its imprints would lie far outside our horizon.  But if we see \textit{some feature} at low~$\ell$, then it should be power suppression.\footnote{Although a power suppression is not unambiguously predicted in this formalism (in the sense of having probability very close to one), the feature is natural, in the sense that we expect extremely large numbers of e-foldings to be unlikely, which implies that the probability for onset in the observable range of comoving scales is not very small~\cite{Bousso:2013uia}.}    For example, a power enhancement at low~$\ell$ would have no motivation in a landscape scenario, and its observation would place considerable pressure on the scenario.  Conversely, the discovery of a power suppression, if confirmed, means that the scenario passed a nontrivial check.

The above discussion becomes more concrete when we compare false vacuum decay to other models for the initial conditions for slow-roll inflation. Consider a class of theories where $\Omega_K=0$ and the field is taken to start out homogeneously somewhere on the potential.  Any such theory is automatically at a disadvantage since, unlike bubble nucleation, it cannot explain why the field began that way, so to some extent this is comparing apples and oranges. Nonetheless, let us allow this comparison.\footnote{Perhaps the most natural way to obtain these initial conditions is from chaotic slow-roll eternal inflation,  although we argued in \cite{Bousso:2013uia} that this is probably less common in the landscape than false vacuum eternal inflation.
}  

Our question then should be whether or not seeing a low-$\ell$ power suppression should cause us to significantly modify our priors for these two types of theories.  We can argue that it should: in the theory with $\Omega_K=0$ there is no particular reason to expect a steepening perturbation at all. For example, it is sometimes argued that ``simple'' potentials like $m^2 \phi^2$ should be expected for large field models, but in these power-law models the scalar power very gradually \textit{increases} as we go to larger and larger scales and there is never any type of sharp feature.\footnote{In stringy constructions of large-field models by contrast, there are typically constraints which prevent the realization of arbitrary large numbers of e-foldings \cite{McAllister:2008hb,Silverstein:2013wua}.  Whether or not the number is small enough to get observable steepening or curvature is model-dependent, and has not been studied in detail.}  

Nonetheless, we can also compare our scenario to models where there is a steepening feature somewhere in the potential, but not a large landscape. In this case detecting steepening leads to a new coincidence problem. Without a statistical distribution of inflaton potentials and anthropic selection along the lines we argued in \cite{Bousso:2013uia}, there is no particular reason to expect this feature to appear between our horizon and the galactic scale. 

The discovery of a steepening feature does not \textit{prove} that our universe arose from the decay of a false vacuum in a large landscape. But it leaves us with one more feature of our universe that has a straightforward explanation in this setting, and which would not otherwise have been expected.

\paragraph{Acknowledgments} We would like to thank Nima Arkani-Hamed, Juan Maldacena, Steve Shenker, Eva Silverstein, Leonard Susskind, and Matias Zaldarriaga for helpful discussions.  The work of R.B.\ is supported by the Berkeley Center for Theoretical Physics, by the National Science Foundation (grant number 1214644), by the Foundational Questions Institute, by ``New Frontiers in Astronomy and Cosmology'', and by the U.S.\ Department of Energy (DE-AC02-05CH11231). L.S. is supported by DOE Early Career Award DE-FG02-12ER41854 and by NSF grant PHY-1068380. D.H. is supported by the Princeton Center for Theoretical Science.

\appendix
\section*{Appendix}
\section{Decorrelation of Gaussian Variables}\label{decapp}
 In this appendix we give a toy model for how one might decorrelate the temperature and the $E$-mode polarization, in the limit where noise is negligible.  First recall that $C_\ell^{TT}$, $C_\ell^{TE}$, and $C_\ell^{EE}$ are defined as two-point functions
\begin{align}\nonumber
\lan a_{\ell m}^{T*} a^T_{\ell',m'}\ran=&C^{TT}_\ell \delta_{\ell \ell'}\delta_{mm'}\\\nonumber
\lan a_{\ell m}^{T*} a^E_{\ell',m'}\ran=&C^{TE}_\ell \delta_{\ell \ell'}\delta_{mm'}\\
\lan a_{\ell m}^{E*} a^E_{\ell',m'}\ran=&C^{EE}_\ell \delta_{\ell \ell'}\delta_{mm'},
\end{align}
where $a^T$ and $a^E$ are the coefficients of the temperature and E-mode polarization maps in appropriate spherical harmonics. Since $\frac{C_\ell^{TE}}{\sqrt{C_\ell^{TT} C_\ell^{EE}}}\neq \pm1$, the $E$ and $T$ modes are not 100\% correlated. For $\ell\gtrsim 10$, the source of this incomplete correlation does not come from very high wavenumbers, but it is just due to order one differences in the visibility functions. We can simply model this situation by two random variables $X$ and $Y$, whose expectation values are
\begin{align}\nonumber
\lan X\ran=\lan Y\ran&=0\\\nonumber
\lan X^2\ran=\lan Y^2\ran&=A\\
\lan XY\ran&=\alpha A.
\end{align}
with $-1\leq \alpha\leq 1$.
Here the quantity $\alpha$ is analogous to $\frac{C_\ell^{TE}}{\sqrt{C_\ell^{TT} C_\ell^{EE}}}$, which is of order $0.5$ in the vicinity of $\ell=30$.   We have rescaled $Y$ so that its variance is equal that of $X$. The goal is then to determine $A$ from measurements of $X$ and $Y$, which is analogous to determining the primordial power spectrum from measuring the temperature and polarization of the CMB.  First consider the situation where we are able to measure only $X$. The obvious thing to do is measure $X^2$, since its expectation value should be $A$.  The variance of this measurement is
\be
\lan (X^2-A)^2\ran=\lan X^4\ran-A^2=2 A^2,
\ee
where we have assumed that $X$ is Gaussian in the sense that higher point functions can be computed by Wick contraction.  We'd now like to see how much including the ability to measure $Y$ decreases the variance on our determination of $A$.  If we had $\alpha=0$ so that $X$ and $Y$ were uncorrelated, then we could clearly just measure $\frac{X^2+Y^2}{2}$, which would have a variance of only $A^2$ and thus would be better by a factor of $\frac{1}{\sqrt{2}}$ in the standard deviation.  Our point here is that in fact we can achieve this improvement for any  $-1\leq\alpha<1$.  The idea is to find an optimal estimator
\be
\hat{A}\equiv c_1 X^2+c_2 Y^2+c_3 XY
\ee
with the dual properties that $\lan \hat{A}\ran=A$ and that $c_1$, $c_2$, and $c_3$ are chosen to minimize the variance
\be
\Delta^2\equiv \lan (\hat{A}-A)^2\ran.
\ee
This is a straightforward Lagrange multiplier problem, which is solved by
\begin{align}\nonumber 
c_1=c_2&=\frac{1}{2(1-\alpha^2)}\\
c_3&=-\frac{\alpha}{1-\alpha^2}.\label{cs}
\end{align}
The variance for this estimator is indeed
\be
\Delta^2=A^2,
\ee
so we have successfully decorrelated the variables $X$ and $Y$.  The results \eqref{cs} are singular in the limit  $\alpha \to \pm1$, but this is to be expected since in that limit $X$ and $Y$ are identical as random variables.  When $\alpha$ is very close to one this procedure will be rather sensitive to small noise. Fortunately we have already noted that $\alpha\approx 0.5$ in the CMB, so decorrelation should be possible.

\bibliographystyle{jhep}
\bibliography{bibliography}

\providecommand{\href}[2]{#2}\begingroup\raggedright\begin{thebibliography}{10}

\bibitem{Linde:1984ir}
A.~D. Linde, {\it {The Inflationary Universe}},  {\em Rept.Prog.Phys.} {\bf 47}
  (1984) 925--986.

\bibitem{Banks:1984cw}
T.~Banks, {\it {T C P, Quantum Gravity, the Cosmological Constant and All
  That...}},  {\em Nucl.Phys.} {\bf B249} (1985) 332.

\bibitem{Weinberg:1987dv}
S.~Weinberg, {\it {Anthropic Bound on the Cosmological Constant}},  {\em
  Phys.Rev.Lett.} {\bf 59} (1987) 2607.

\bibitem{Bousso:2000xa}
R.~Bousso and J.~Polchinski, {\it {Quantization of four form fluxes and
  dynamical neutralization of the cosmological constant}},  {\em JHEP} {\bf
  0006} (2000) 006, [\href{http://xxx.lanl.gov/abs/hep-th/0004134}{{\tt
  hep-th/0004134}}].

\bibitem{Kachru:2003aw}
S.~Kachru, R.~Kallosh, A.~D. Linde, and S.~P. Trivedi, {\it {De Sitter vacua in
  string theory}},  {\em Phys.Rev.} {\bf D68} (2003) 046005,
  [\href{http://xxx.lanl.gov/abs/hep-th/0301240}{{\tt hep-th/0301240}}].

\bibitem{Susskind:2003kw}
L.~Susskind, {\it {The Anthropic landscape of string theory}},
  \href{http://xxx.lanl.gov/abs/hep-th/0302219}{{\tt hep-th/0302219}}.

\bibitem{Freivogel:2005vv}
B.~Freivogel, M.~Kleban, M.~Rodriguez~Martinez, and L.~Susskind, {\it
  {Observational consequences of a landscape}},  {\em JHEP} {\bf 0603} (2006)
  039, [\href{http://xxx.lanl.gov/abs/hep-th/0505232}{{\tt hep-th/0505232}}].

\bibitem{Bousso:2013uia}
R.~Bousso, D.~Harlow, and L.~Senatore, {\it {Inflation after False Vacuum
  Decay: Observational Prospects after Planck}},
  \href{http://xxx.lanl.gov/abs/1309.4060}{{\tt arXiv:1309.4060}}.

\bibitem{Ade:2014xna}
{\bf BICEP2 Collaboration} Collaboration, P.~Ade et~al., {\it {BICEP2 I:
  Detection Of B-mode Polarization at Degree Angular Scales}},
  \href{http://xxx.lanl.gov/abs/1403.3985}{{\tt arXiv:1403.3985}}.

\bibitem{Ade:2014gua}
{\bf BICEP2 Collaboration} Collaboration, P.~A.~R. Ade et~al., {\it {BICEP2 II:
  Experiment and Three-Year Data Set}},
  \href{http://xxx.lanl.gov/abs/1403.4302}{{\tt arXiv:1403.4302}}.

\bibitem{Senatore:2011sp}
L.~Senatore, E.~Silverstein, and M.~Zaldarriaga, {\it {New Sources of
  Gravitational Waves during Inflation}},
  \href{http://xxx.lanl.gov/abs/1109.0542}{{\tt arXiv:1109.0542}}.

\bibitem{Ade:2013uln}
{\bf Planck Collaboration} Collaboration, P.~Ade et~al., {\it {Planck 2013
  results. XXII. Constraints on inflation}},
  \href{http://xxx.lanl.gov/abs/1303.5082}{{\tt arXiv:1303.5082}}.

\bibitem{Easther:2006tv}
R.~Easther and H.~Peiris, {\it {Implications of a Running Spectral Index for
  Slow Roll Inflation}},  {\em JCAP} {\bf 0609} (2006) 010,
  [\href{http://xxx.lanl.gov/abs/astro-ph/0604214}{{\tt astro-ph/0604214}}].

\bibitem{Miranda:2014wga}
V.~Miranda, W.~Hu, and P.~Adshead, {\it {Steps to Reconcile Inflationary Tensor
  and Scalar Spectra}},  \href{http://xxx.lanl.gov/abs/1403.5231}{{\tt
  arXiv:1403.5231}}.

\bibitem{Smith:2014kka}
K.~M. Smith, C.~Dvorkin, L.~Boyle, N.~Turok, M.~Halpern, et~al., {\it {On
  quantifying and resolving the BICEP2/Planck tension over gravitational
  waves}},  \href{http://xxx.lanl.gov/abs/1404.0373}{{\tt arXiv:1404.0373}}.

\bibitem{Hazra:2014aea}
D.~K. Hazra, A.~Shafieloo, G.~F. Smoot, and A.~A. Starobinsky, {\it {Ruling out
  the power-law form of the scalar primordial spectrum}},
  \href{http://xxx.lanl.gov/abs/1403.7786}{{\tt arXiv:1403.7786}}.

\bibitem{Hazra:2014jka}
D.~K. Hazra, A.~Shafieloo, and G.~F. Smoot, {\it {Whipped inflation}},
  \href{http://xxx.lanl.gov/abs/1404.0360}{{\tt arXiv:1404.0360}}.

\bibitem{Linde:1998iw}
A.~D. Linde, {\it {A Toy model for open inflation}},  {\em Phys.Rev.} {\bf D59}
  (1999) 023503, [\href{http://xxx.lanl.gov/abs/hep-ph/9807493}{{\tt
  hep-ph/9807493}}].

\bibitem{Garriga:1998he}
J.~Garriga, X.~Montes, M.~Sasaki, and T.~Tanaka, {\it {Spectrum of cosmological
  perturbations in the one bubble open universe}},  {\em Nucl.Phys.} {\bf B551}
  (1999) 317--373, [\href{http://xxx.lanl.gov/abs/astro-ph/9811257}{{\tt
  astro-ph/9811257}}].

\bibitem{Linde:1999wv}
A.~D. Linde, M.~Sasaki, and T.~Tanaka, {\it {CMB in open inflation}},  {\em
  Phys.Rev.} {\bf D59} (1999) 123522,
  [\href{http://xxx.lanl.gov/abs/astro-ph/9901135}{{\tt astro-ph/9901135}}].

\bibitem{Contaldi:2003zv}
C.~R. Contaldi, M.~Peloso, L.~Kofman, and A.~D. Linde, {\it {Suppressing the
  lower multipoles in the CMB anisotropies}},  {\em JCAP} {\bf 0307} (2003)
  002, [\href{http://xxx.lanl.gov/abs/astro-ph/0303636}{{\tt
  astro-ph/0303636}}].

\bibitem{Yamauchi:2011qq}
D.~Yamauchi, A.~Linde, A.~Naruko, M.~Sasaki, and T.~Tanaka, {\it {Open
  inflation in the landscape}},  {\em Phys.Rev.} {\bf D84} (2011) 043513,
  [\href{http://xxx.lanl.gov/abs/1105.2674}{{\tt arXiv:1105.2674}}].

\bibitem{Liddle:2013czu}
A.~R. Liddle and M.~Cortês, {\it {Cosmic microwave background anomalies in an
  open universe}},  {\em Phys.Rev.Lett.} {\bf 111} (2013) 111302,
  [\href{http://xxx.lanl.gov/abs/1306.5698}{{\tt arXiv:1306.5698}}].

\bibitem{Dudas:2012vv}
E.~Dudas, N.~Kitazawa, S.~Patil, and A.~Sagnotti, {\it {CMB Imprints of a
  Pre-Inflationary Climbing Phase}},  {\em JCAP} {\bf 1205} (2012) 012,
  [\href{http://xxx.lanl.gov/abs/1202.6630}{{\tt arXiv:1202.6630}}].

\bibitem{Pedro:2013pba}
F.~G. Pedro and A.~Westphal, {\it {Low-l CMB Power Loss in String Inflation}},
  \href{http://xxx.lanl.gov/abs/1309.3413}{{\tt arXiv:1309.3413}}.

\bibitem{Weinberg:2008zzc}
S.~Weinberg, {\it {Cosmology}}, .

\bibitem{Lesgourgues:2011re}
J.~Lesgourgues, {\it {The Cosmic Linear Anisotropy Solving System (CLASS) I:
  Overview}},  \href{http://xxx.lanl.gov/abs/1104.2932}{{\tt arXiv:1104.2932}}.

\bibitem{Blas:2011rf}
D.~Blas, J.~Lesgourgues, and T.~Tram, {\it {The Cosmic Linear Anisotropy
  Solving System (CLASS) II: Approximation schemes}},  {\em JCAP} {\bf 1107}
  (2011) 034, [\href{http://xxx.lanl.gov/abs/1104.2933}{{\tt
  arXiv:1104.2933}}].

\bibitem{Ade:2013kta}
{\bf Planck Collaboration} Collaboration, P.~Ade et~al., {\it {Planck 2013
  results. XV. CMB power spectra and likelihood}},
  \href{http://xxx.lanl.gov/abs/1303.5075}{{\tt arXiv:1303.5075}}.

\bibitem{Ade:2013zuv}
{\bf Planck Collaboration} Collaboration, P.~Ade et~al., {\it {Planck 2013
  results. XVI. Cosmological parameters}},
  \href{http://xxx.lanl.gov/abs/1303.5076}{{\tt arXiv:1303.5076}}.

\bibitem{Spergel:2013rxa}
D.~Spergel, R.~Flauger, and R.~Hlozek, {\it {Planck Data Reconsidered}},
  \href{http://xxx.lanl.gov/abs/1312.3313}{{\tt arXiv:1312.3313}}.

\bibitem{Hinshaw:2012aka}
{\bf WMAP} Collaboration, G.~Hinshaw et~al., {\it {Nine-Year Wilkinson
  Microwave Anisotropy Probe (WMAP) Observations: Cosmological Parameter
  Results}},  {\em Astrophys.J.Suppl.} {\bf 208} (2013) 19,
  [\href{http://xxx.lanl.gov/abs/1212.5226}{{\tt arXiv:1212.5226}}].

\bibitem{Mortonson:2009qv}
M.~J. Mortonson, C.~Dvorkin, H.~V. Peiris, and W.~Hu, {\it {CMB polarization
  features from inflation versus reionization}},  {\em Phys.Rev.} {\bf D79}
  (2009) 103519, [\href{http://xxx.lanl.gov/abs/0903.4920}{{\tt
  arXiv:0903.4920}}].

\bibitem{Kleban:2012ph}
M.~Kleban and M.~Schillo, {\it {Spatial Curvature Falsifies Eternal
  Inflation}},  {\em JCAP} {\bf 1206} (2012) 029,
  [\href{http://xxx.lanl.gov/abs/1202.5037}{{\tt arXiv:1202.5037}}].

\bibitem{McAllister:2008hb}
L.~McAllister, E.~Silverstein, and A.~Westphal, {\it {Gravity Waves and Linear
  Inflation from Axion Monodromy}},  {\em Phys.Rev.} {\bf D82} (2010) 046003,
  [\href{http://xxx.lanl.gov/abs/0808.0706}{{\tt arXiv:0808.0706}}].

\bibitem{Silverstein:2013wua}
E.~Silverstein, {\it {Les Houches lectures on inflationary observables and
  string theory}},  \href{http://xxx.lanl.gov/abs/1311.2312}{{\tt
  arXiv:1311.2312}}.

\end{thebibliography}\endgroup

\end{document}